# Report for the Edinburgh Tram Inquiry

**Prepared by**
Prof. Bent Flyvbjerg* and Dr. Alexander Budzier

*Lead author; all opinions expressed in this report are the opinions of the lead author and he accepts responsibility for all errors and omissions.

Version 14

February 2018

# Table of Contents

# Glossary

| Term | Definition |
| --- | --- |
| BBS | Bilfinger Berger Siemens consortium |
| CAF | Construcciones y Auxiliar de Ferrocarriles SA |
| CEC | City of Edinburgh Council |
| DBFMO | Design, build, finance, manage and operate contracts |
| DfT | Department for Transport |
| GRIP | Guide to Rail Investment Process |
| HMT | HM Treasury |
| INFRACO | Infrastructure Construction Contract |
| NAO | National Audit Office |
| OB | Optimism Bias |
| OGC | Office of Government Commerce |
| SDS | System Design Service Contract |
| STAG | Scottish Transport Appraisal Guidance |
| TAG | Transport Appraisal Guidance |
| TPB | Tram Project Board |
| TEL | Transport Edinburgh Ltd |
| TIE | Transport Initiatives Edinburgh |
| TRAMCO | Tram Vehicle Contract |

# Executive Summary

This report reviewed the Edinburgh tram project's risk management. Projects frequently overrun their cost and timelines and fall short of the intended benefits. Cost, schedule and benefit risk of projects need to be carefully considered to avoid this.

Often project risks are underreported or ignored. Underreporting may be intentional or not. Non-intentional underreporting is caused by optimism bias with planners. Intentional underreporting is caused by strategic misrepresentation. Optimism bias and strategic misrepresentation are root causes of inaccurate assessments of project risk.

Probabilistic forecasts are current best practice of presenting risk estimates and allowing decision makers to appraise projects given different required levels of certainty of estimates. Risk appetite typically differs for the three key questions asked during project appraisal:
- Is the project economically viable?
- Is the project affordable?
- What should the cost, schedule, and benefit targets be at different levels?

Risks are typically assessed using the inside view, which is subject to optimism and strategic misrepresentation. The cure is to de-bias project estimates by taking an outside view. Reference class forecasting, including optimism bias uplifts, is a well-established method to systematically take the outside view.

The report reviewed the guidance available when the Edinburgh tram project produced the Draft Interim Outline Business Case (May 2005) and the Final Business Case (December 2007).

In the experts' judgement, the approach taken to estimates, risk and optimism bias in the Edinburgh tram project was generally similar to the approach of other projects of a similar nature at the time. Equally, the mitigation measures planned and the work to understand risk were similar to those of other projects.

In the view of the experts, the Draft Interim Outline Business Case is optimistic with regards to cost risk. The project team argued that it would deliver according to the budget envelope with more than 95% certainty while other data, which were available to the planners, suggested that a 20% risk of exceeding the funding envelope existed, i.e. four times the estimate.

In the experts' view, there are some doubts about the quality of the quantitative risk analysis which estimated a cost risk of 15% at P90, which seems low given the high level of confidence and the evidence in the official guidance documents available to the project at the time. Optimism bias is likely to have entered the risk assessment process during quantitative risk analysis, which is understandable, because quantitative risk analysis is based on expert judgement and such judgement has been proven to be prone to optimism bias. Instead of reducing optimism bias, quantitative risk analysis seems to have increased this for the Edinburgh Tram. In addition, the optimism was perpetuated by portraying the cost estimate with a high degree of confidence.

# 1. Key Concepts

## Risk

Most projects change during the project cycle from idea to reality. Changes may be due to uncertainty regarding the level of ambition, the exact corridor, the technical standards, safety, environment, project interfaces, geotechnical conditions, etc. In addition, the prices and quantities of project components are subject to uncertainty and change.

Risk is conventionally regarded as the adverse consequence of change with a consequence that projects fail to meet deadlines and cost targets (Smith et al. 2006).

In terms of risk, most appraisals of projects assume, or pretend to assume, that infrastructure projects exist in a world where things go according to plan. In reality, the world of project preparation and implementation is a highly risky one where things happen only with a certain probability and rarely turn out as originally intended.

Nine out of ten such projects have cost overruns; overruns of up to 50% in real terms are common, over 50% are not uncommon. The cost overrun for the Channel Tunnel, the longest underwater rail tunnel in Europe, connecting the United Kingdom and France, was 80% in real terms. The cost overruns for the Denver International Airport were 200% and for Boston's Big Dig, 220%.

Overrun is a problem in private as well as public sector projects, and things are not improving; overruns have stayed high and constant for the 70-year period for which comparable data exist. Geography doesn't seem to matter either; all countries and continents for which data are available suffer from overruns. Similarly, benefit shortfalls of up to 50% are also common and above 50% not uncommon, again with no signs of improvements over time and geography (Flyvbjerg, Holm, & Buhl 2002).

Hence, some degree of risk for cost overrun, schedule delay and benefit shortfall will always exist and is important to consider for project appraisal, programming, budget setting and project cost control. Risk is however not unknown and should be duly reflected in the project documentation at any given stage.

## Optimism Bias and Strategic Misrepresentation

Often project risks are underreported or ignored. Underreporting may be intentional or not. Non-intentional underreporting is caused by optimism bias with planners. Intentional underreporting is caused by strategic misrepresentation. Optimism bias and strategic misrepresentation are root causes of inaccurate assessments of project risk.

Psychologists tend to explain underreporting of risk in terms of optimism bias, that is, a cognitive predisposition found with most people to judge future events in a more positive light than is warranted by actual experience. Clearly an optimistic budget is a low budget, and cost overrun follows.

Economists and political scientists tend to explain underreporting of budget risk in terms of strategic misrepresentation, or political bias. Here, when forecasting the outcomes of projects, forecasters and planners deliberately and strategically overestimate benefits and underestimate costs in order to increase the likelihood that it is their projects, and not the competition's, that gain approval and funding.

Political bias can be traced to political and organizational pressures, for instance competition for scarce funds or jockeying for position, and to lack of incentive alignment. Optimism bias and political bias are both deception, but where the latter is deliberate, the former is not. Optimism bias is self-deception.

*Figure 1 Explanations of Risk in Projects: Optimism Bias and Strategic Misrepresentation*

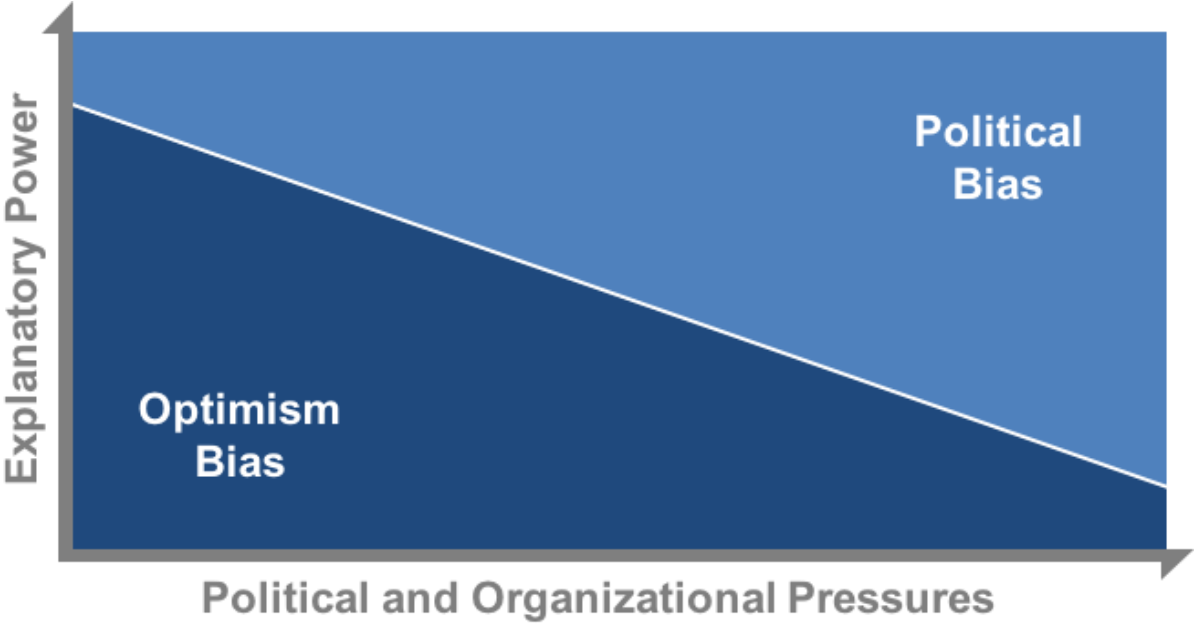

Although the two types of explanation are different, the result is the same: inaccurate forecasts and inflated benefit-cost ratios, which realise as cost overrun and benefit shortfall.

As illustrated schematically in Figure 1, explanations of optimism bias have their relative merit in situations where political and organizational pressures are absent or low, whereas such explanations hold less power in situations where political pressures are high.

Conversely, explanations of strategic misrepresentation have their relative merit where political and organizational pressures are high, while they become less relevant when such pressures are not present. Thus, rather than compete, the two types of explanation complement each other: one is strong where the other is weak, and both explanations are necessary to understand the pervasiveness of inaccuracy and risk in project planning—and how to curb it.

# 2. Probability and Risk

A probabilistic forecast presents the distributional information of the forecast. For example, a probabilistic cost risk forecast presents the forecasted risk not as a single point estimate but as a range of outcomes given a range of likelihoods. Conventionally, the simplest form of a probabilistic forecast is a forecast for the best case, most likely case and the worst case.

Better probabilistic forecasts represent the full distribution of forecasted outcomes. This is often reported as shown in Figure 2, which shows the probability of the risk on the x-axis and the probability of the cost risk on the y-axis. The probability is commonly presented as the certainty of the estimate. For example, P50 means that the forecast is 50% certain and has thus a 50% likelihood of being exceeded, P80 means that the forecast is 80% certain and has a 20% likelihood of being exceeded. The mean of the distribution is in some guidance (e.g. WebTAG) labelled P(Mean).

*Figure 2 Conceptual Probabilistic Cost Risk Forecast*

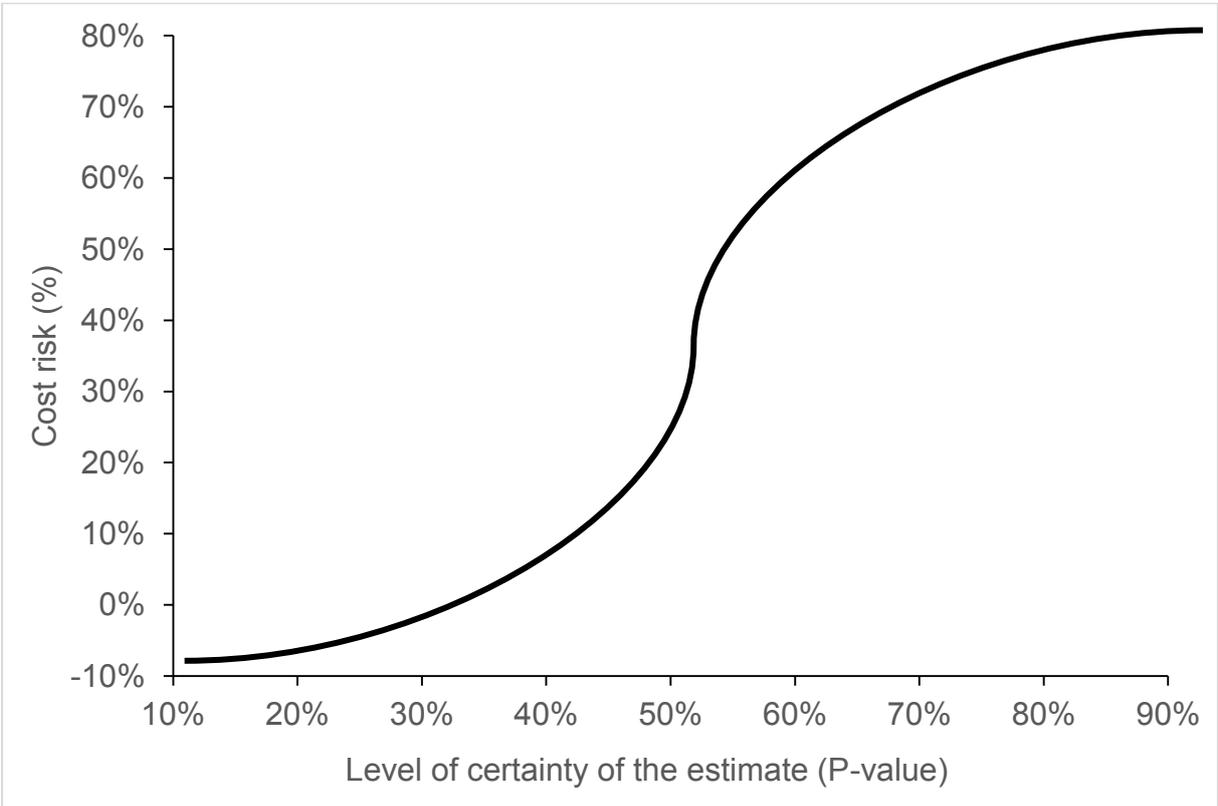

Figure 2 depicts and idealized S-curve, which implies that the risks are normally and symmetrically distributed around the median (P50). In reality risks are rarely normally and symmetrically distributed. Figure 3 shows the cost risk of light rail systems. The risk forecast is based on the analysis of data from 63 historic light rail projects and follows the methodology used in Flyvbjerg and COWI (2004). Figure 3 differs from the idealized S-curve in Figure 2, in particular for the tail of the distribution which for estimates above P60 is convex in Figure 3 and not concave like in Figure 2. These tail risks, i.e. low likelihood risks with very high cost overruns are particularly important.

Observations that form these tails are projects with very high cost overruns. These projects are also called "Black Swans", a popular term for extreme events with massively negative outcomes (Taleb, 2010). The statistical properties of Black Swans affect the ability to forecast, because Black Swans are not expected to occur in a well-defined, deterministic Newtonian world of cause, effect, and control. In statistical terms, Black Swans are outliers. Outliers are commonly defined to be 1.5 inter-quartile ranges (the difference between the top and bottom quartile) away from the top quartile. Defined in this manner, in the data of light rail projects outliers are projects with cost overruns ≥ 143%. 6% of the observations in the reference class are classified as outliers. The cost overrun, in real terms, of the Edinburgh Tram was +52%. Thus the Edinburgh Tram was not a Black Swan. A common misconception is that Black Swans are freak occurrences to be excluded from reference classes. However, managers should not ignore Black Swans because Black-Swan projects are not caused by catastrophic risks materializing (e.g. disease outbreaks, terrorism) but are typically the result of multiple adverse events occurring simultaneously. Thus, while they cannot be predicted managers can learn from them to reduce their projects' exposure to Black Swans.

*Figure 3 Probabilistic Forecast of Cost Risk of Light Rail Projects (n=63; Source: Authors' Database)*

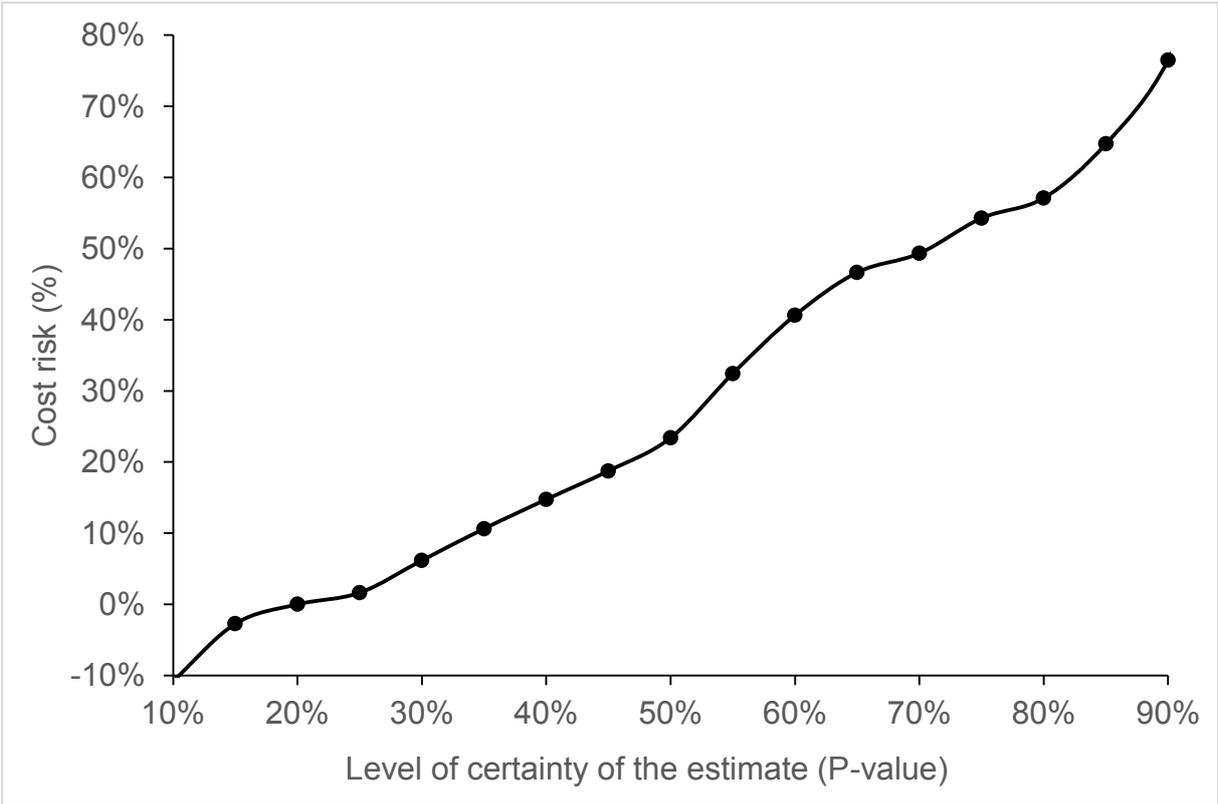

# 3. Assessment of Risk

Conventionally, risk is estimated by identifying events that have an impact on the estimates of quantities or prices or both (AXELOS, 2011). Risk commonly identified in expert workshops cataloguing possible events that might happen and impact the project. The identified risks are logged in a risk register, where the likelihood and impact of each risk is assessed and mitigating actions are tracked.

The impacts and the likelihood are quantified, typically using subjective best guesses and rarely based on hard empirical data from past projects. The total value of the risk register is then summed up as the risk estimate. This, in its simplest form, is the weighted average of likelihood multiplied by the impact of all identified risks.

Currently, the industry standard is to use a Monte Carlo simulation to model the full range of futures, from no risks happening to all identified risks happening to a project. Monte Carlo simulations can also account for correlations between risks, e.g. if the occurrence of one risk event increases or decreases the likelihood of a related risk.

The result of such risk assessment combined with Monte Carlo simulations looks objective and quantitative, even scientific. In reality it is subjective and qualitative, based on judgement.

Conventional cost risk estimation takes an inside view of a project: planners look at all the constituent elements of a project that need to be completed and the uncertainty in the cost estimate of each element and the cost impact of adverse events that might impact each element make up the risk estimate.

The inside view has been shown to result in optimistic forecasts (Kahneman and Lovallo 2003), i.e. to produce a systematic underreporting of the level of project risk.

Kahneman and Tversky's (1979a, b) found that human judgment is generally optimistic due to overconfidence and insufficient regard for distributional information about outcomes.

Thus, people will underestimate the costs, completion times, and risks of planned actions, whereas they will overestimate the benefits of the same actions. Such errors of judgment are shared by experts and laypeople alike, according to Kahneman and Tversky.

Lovallo and Kahneman (2003: 58) call such common behaviour the "planning fallacy." They argue that this fallacy stems from actors taking an "inside view" focusing on the constituents of the specific planned action rather than on the outcomes of similar actions that have already been completed.

Research into the track record of past estimates (e.g. Flyvbjerg and COWI 2004, Flyvbjerg 2014, 2016) shows that project cost estimates are systematically and consistently lower than the actual outturn cost. The data show that inside-view cost estimates are biased, i.e. systematically underestimating cost risk.

# 4. Assessment of Optimism Bias

The cure for the inside view is to take the so-called "outside view". The outside view pools lessons from past projects and uses these as distributional information to statistically predict the outcome of another project.

Kahneman and Tversky (1979b) argue that the prevalent tendency to underweight or ignore distributional information is perhaps the major source of error in forecasting. Planners should therefore make every effort to frame the forecasting problem so as to facilitate utilizing all the distributional information that is available.

This may be considered the single most important piece of advice regarding how to increase accuracy in forecasting through improved methods, according to Kahneman (2011).

The outside view makes explicit, empirically based adjustments to cost estimates (Flyvbjerg, Holm, and Buhl 2003; Mott MacDonald 2002; HM Treasury 2003). In order to be accurate, these adjustments should be based on data from past projects or similar projects elsewhere, and adjusted for the unique characteristics of the project in hand.

Reference class forecasting is one method for systematically taking an outside view on planned actions. Reference class forecasting places particular emphasis on relevant distributional information because such information is most significant to the production of accurate forecasts.

Reference class forecasting follows three steps:

1. Identify a sample of past, similar projects – typically a minimum of 20-30 projects are enough to get started, but the more projects the better;
2. Establish the risk of the variable in question based on these projects – e.g. identify the cost overruns of these projects; and
3. Adjust the current estimate – through an uplift or by asking whether the project at hand is more or less risky than projects in the reference class, resulting in an adjusted uplift.

The final step in the reference class forecasting process considers whether the project at hand is more or less risky than the projects in the reference class. It should be stressed that unless this consideration is based on objective evidence, optimism bias might be re-introduced into estimates.

In addition, planners might consider downward adjustments of optimism bias uplifts based on capability assessment, e.g. ability to identify and address risks early or commercial structures established with a goal to transfer risk as much as possible. These assessments are all subjective, because they are untested at the planning stage.

Thus, they ought to be viewed with suspicion. For example, the risk mitigation effect of the Edinburgh Tram project's procurement strategy was untested at the final business case stage and in hindsight might be considered optimistic.

Thus, projects have to face the competing pressures of the funder's demands to deliver as much value for money as possible and being prudent in their risk assessment. In practice, downward adjustments to risk and optimism bias uplifts ought to pass a critical test of objectivity to be justified.

Risk assessments based on the inside and outside views follow a similar process of (1) identification, (2) quantification, and (3) management of risks. Both risk assessment approaches result in a cost estimate that depends on the desired level of certainty of the cost estimate.

The difference between inside and outside view approaches to forecasting is that the former are based on subjective and thus biased judgement and the latter are based on objective data, thus circumventing optimism bias and the planning fallacy.

In theory, if risk assessment based on the inside view is done without bias the risk estimate should either match the outside view estimate or give very robust, reliable and clear evidence why risks are higher or lower than the outside view's assessment of risks.

Reference Class Forecasts of project cost risk were used to establish the Optimism Bias Uplifts that are required by DfT's Transport Appraisal Guidance (2006) and the HMT Greenbook (2003). The guidance recommends to uplift project cost estimates by adding the forecasted cost risk to the cost estimate, thus de-biasing the cost estimate.

In the case of transport the categories for which data is provided by DfT are road, rail, and fixed link, building, IT, standard civil and non-standard civil projects (Flyvbjerg and COWI 2004 and Mott MacDonald 2002).

# 6. Summary of Guidance Documents on Optimism Bias

## Mott MacDonald's Review

The review had the express purpose to provide an evidence base for the HMT Green Book, which was then under development. Thus, the review attempted to determine measures for the optimism bias in UK large public procurements; identified project risk areas; and provided "guidance for managing project risk areas through the application of best practices to minimise optimism in project estimates" (Mott MacDonald 2002, p. S-1).

The review urges that optimism bias ought to be considered for cost, duration, and benefits estimates (p. S-2). The review argues that optimism, which manifests in cost overruns, delays and benefits shortfalls, is caused by optimistic assessment of risks. Thus, a project ought to assess "the degree to which there is evidence that project risks have been identified and will be managed…" (p. S-2).

The review collected data on 50 completed projects. For non-standard civils (such as the Edinburgh Tram project) 13 projects were included in the data.

In terms of concrete guidance, the review makes the following recommendations:

- Establish a clear governance process, including stage gates, clearly defined objectives, use of project reviews, full life cycle view of the project, benefits planning and measurement, risk and value management, identified project sponsor;
- Risk management during the appraisal process; incl.
    o Use of experienced appraisers;
    o Full analysis of all risks;
    o Understanding of the risks at the specific stage in the project's life cycle;
    o Strong business case; and
    o Balance the cost of risk management with the risk exposure of the project.
- Calculate the optimism bias in the project, through:
    o Identify the project type (e.g. non-standard civils);
    o Use the upper bound value as a starting point (e.g. 66% for CAPEX and 25% for works duration for non-standard civils);
    o Reduce the upper bound by the value that project risks are being managed based on clear and tangible evidence that those risks are managed. For non-standard civils, the optimism bias for CAPEX is broken down as follows:
        ▪ Procurement (2%);
        ▪ Design complexity (8%);
        ▪ Innovation (9%);
        ▪ Environmental impact (5%);
        ▪ Inadequacy of the business case (35%);
        ▪ Funding availability (5%);
        ▪ Project management team (2%);

- Poor project intelligence (9%);
- Site characteristics (5%);
- Economic influences (3%);
- Legislation and regulation (8%);
- Technology (8%); and
- Other external influences (1%).
  - Appraise the project given the calculated level of optimism bias; let only projects with low optimism bias continue to the next stage gate.

## HMT Green Book

The HM Treasury Green Book puts the findings of the Mott MacDonald review into practice.

With regards to optimism bias the Green Book states: "To redress this [optimistic] tendency, appraisers should make explicit adjustments for this bias. These will take the form of increasing estimates of the costs and decreasing, and delaying the receipt of, estimated benefits. Sensitivity analysis should be used to test assumptions about operating costs and expected benefits." (p. 29) and further "Adjustments should be empirically based, (e.g. using data from past projects or similar projects elsewhere), and adjusted for the unique characteristics of the project in hand. Cross-departmental guidance for generic project categories is available, and should be used in the absence of more specific evidence" (p.29).

With regards to conventional risk appraisal the Green Book states: "It is good practice to add a risk premium to provide the full expected value of the Base Case. […] in the early stages of an appraisal, this risk premium may be encompassed by a general uplift to a project's net present value, to offset and adjust for undue optimism. But as appraisal proceeds, more project specific risks will have been identified, thus reducing the need for the more general optimism bias" (p.29).

Specifically, the Green Book outlines three steps to deal with optimism bias:

1. "Estimate the capital costs of each option,
2. Apply adjustments to these estimates, based on the best available empirical evidence relevant to the stage of the appraisal; and,
3. Subsequently, reduce these adjustments according to the extent of confidence in the capital costs' estimates, the extent of management of generic risks, and the extent of work undertaken to identify and mitigate project specific risk." (p. 85)

With regards to risk management, the Green Book recommends to establish organisational-level risk management (i.e. establishing a risk management framework, senior management support, communication of policies, and embedding processes). In addition, at the project level the Green Book specifically recommends the use of conventional sensitivity analysis, scenarios, and Monte Carlo analysis at the appraisal stages, and risk registers during execution of projects.

# DfT's Guidance on Procedures for Dealing with Optimism Bias

The DfT Guidance builds on the HMT Green Book and the Mott MacDonald study with additional evidence for transport projects.

The Guidance is based on a sample of 260 transport projects, which includes a reference class of 46 rail projects (i.e. the reference class applicable to the Edinburgh tram project). The uplifts recommended for rail projects, including light rail, are P50 = 40%, P80 = 57%. The analysis warns that "It may be argued that uplifts should be adjusted downward as risk assessment and management improves over time and risks are thus mitigated. It is however our view, that planners and forecasters should carry out such downward adjustment of uplifts only when warranted by firm empirical evidence." (p. 34)

In addition, to the optimism bias uplifts the analysis recommends to establish realistic budgeting, introduce financial incentives against cost overruns, and formalise requirements for high-quality cost-risk assessment at the business case stage, and independent appraisals. Risk assessment should include generic risk analysis checklists, risk identification workshops, statistical scenario analysis, and assessments of market structure and levels of competition.

# STAG

The Scottish Transport Appraisal Guidance issued in September 2003 reiterates the recommendations from the Mott MacDonald report and the HMT Green Book. It recommends the use of the optimism bias uplifts established in these documents and "…as appraisal proceeds, more project-specific risks will have been identified thus reducing the need for the application of more general optimism bias factors." (p. 12-4) For this, STAG recommends conventional quantitative risk analysis (risk = impact * likelihood) or Monte Carlo analysis.

STAG states that only when all risks are correctly identified will there be no need for optimism bias uplifts. However, since unanticipated risks remain even in well-developed projects, a residual optimism bias uplift should be allocated in form of a contingency. STAG stipulates that this residual optimism should be valued at the lower bounds of values described in the Mott MacDonald report (i.e. 6% on CAPEX and 3% on works duration for non-standard civils).

Lastly, to assess the business case STAG recommends the use of scenario analyses.

# DfT's TAG

The DfT's Transport Appraisal Guidance sets out the importance of cost estimation to appraise project proposals and options. The three key elements of a cost estimate are:
- Base cost, which includes inflation allowance;
- Risk adjustment, which covers all identified and quantified risks; and
- Optimism bias adjustment, which is an uplift applied after risk adjustments have been made to the base cost estimate.

For the risk adjustment, the department requires a quantitative risk assessment for all projects with cost greater than £5m (§3.2.3). The guidance also retired the use of contingencies for non-quantifiable or difficult to quantify risks (§3.2.8); and guides projects to ignore catastrophic risks (§3.2.9).

The risk assessment ought to reflect the size and stage of the project in hand. In general, the assessment is based on a risk register that is valued by assessing likelihood and impact of risks (§3.3).

The department requires taking the weighted mean value of the risk register and adding this to the base cost as the basis for the optimism bias adjustment (§3.3.17). In addition, Monte Carlo analysis is suggested to model the correlation between risks. Again, the mean estimate of the Monte Carlo analysis ought to be used in the risk assessment.

In addition to the risk assessment, the guidance requires a risk management strategy. The strategy needs to explain how risks are dealt with (e.g. by tolerating, treating, transferring or terminating the risky activity) and how risks are controlled through preventative controls, corrective controls, directive controls or detective controls (§3.2.13-3.2.15).

For the optimism bias adjustment, the TAG follows the HMT Green Book and the DfT's Guidance on Procedures for Dealing with optimism bias. The procedure consists of four steps: (1) identify project type, (2) identify stage of development, (3) apply the recommended uplifts, and (4) provide sensitivity analysis around the central estimate.

For light rail projects those uplifts to be applied to the risk-adjusted scheme cost estimate (§3.7.7) are:
- Stage 1 (GRIP Stage 1: pre-feasibility) = 66%;
- Stage 2 (GRIP Stage 3: option selection) = 40%; and
- Stage 3 (GRIP Stage 5: design development) = 6% (Table 9).

The guidance states that DfT does not expect to see lower adjustments than these stated and if they are reduced the department expects a clearly documented process and an evidence base for any reductions.

Sensitivity analysis is required and the guidance recommends the use of the lower bounds, central case, and upper bounds of the optimism bias adjustments to establish three scenarios for the sensitivity analysis.

The current guidance (TAG Unit 5.3, as of December 2015) on rail appraisals requires that at GRIP stages 1-3 the base cost excluding QRA adjustment is used as the basis for the optimism bias uplift; and at GRIP stages 4-5 base cost plus QRA adjustment are used as the basis for optimism bias uplift (§2.5.3).

Currently, the following levels of risk adjustment and optimism bias adjustment are required (table 3, p. 6), consistent with the TAG guidance in place since September 2006.

*Table 1 Risk Treatment at Different Levels of Project Development*

| Project Development Level (Equivalent to Network Rail's GRIP stages) | Level 1 | Level 2 | Level 3 | Level 4 | Level 5 |
|---|---|---|---|---|---|
| **Activity** | Project Definition | Pre-feasibility | Option Selection | Single Option Refinement | Design Develop-ment |
| **CAPEX QRA, contingency** | No | No | No | QRA at mean estimate | QRA at mean estimate |
| **CAPEX Optimism Bias (% of present value)** | 66% | 50% | 40% | 18% | 6% |
| **OPEX QRA, contingency** | No | No | No | QRA at mean estimate | QRA at mean estimate |
| **OPEX Optimism Bias** | 41% of present value | 1.6% per annum | 1% per annum | 1% per annum | 1% per annum |

# 7. Optimism Bias Treatment at the Edinburgh Tram Project

The final business case for the Edinburgh Tram forecasted the cost of Phase 1a (Airport to Newhaven) at £498m and Phase 1b (Roseburn to Granton) at £87m (§1.65), against a funding commitment of up to £500m from the Scottish Government and £45m from the CEC (§1.73). The final business case estimates that £5m could be saved by concurrent construction of both Phase 1a and Phase 1b (§1.65).

The present value of the cost of Phase 1a was estimated at £335m against benefits of £592m in 2002 prices. This achieved a cost-benefit ratio of 1.77 (§1.97).

99.9% of the base cost estimate, according to the final business case, was based on rates and prices received in bids, with the remainder based on known rates and land valuation (§1.66).

The final business case stated that the cost estimate included a risk adjustment based on "rigorous quantitative risk analysis" (§1.68). The risk adjustment was set at 15% of the construction period base cost (§10.35), which was expected to be a P90 estimate (§11.42).

The final business case compared the cost estimate of Phase 1a (at £498m at P90) against the funding committed (of £545m) and concluded that the project had "a 14% headroom above and beyond the 15% risk allowance" (§1.73).

The final business case explicitly states that TIE is aware of the risks that are retained and that those are managed and priced through a risk register (§1.84). The key risks identified are the utility diversions, scope and specification changes, and obtaining consent and approvals (§1.86).

The cost-benefit analysis of the final business case (§1.97 and §1.98) does not mention a risk uplift for operations costs or revenues. Under current guidance introduced later, this is required – but not when the final business case was being developed.

The final business case also states that "The UK light rail sector has encountered difficulties in the last six years. Those have affected both existing projects and those in procurement. On the earliest schemes, it appears that the private sector showed over-confidence in respect of the risks faced, and in some cases, the public sector showed a lack of foresight." (§7.3) The business case argues that this experience is reflected in the procurement and risk strategy for the Edinburgh Tram. In effect, the project planned to transfer design, construction and maintenance performance risks to the private sector; minimise the risk premium in bids; mitigate utility diversion risk; and gain early contractor involvement (§7.9).

As a result, the procurement strategy planned to award the SDS contract on a fixed-price lump-sum basis. The MUDFA is a re-measurement contract with benefits-sharing of cost savings. The INFRACO is a lump sum tied into milestones. The procurement strategy planned for an option to provide full novation of the SDS contract to the INFRACO contract after contract award of the latter. This, the business case argues, will remove the interface risk between design and construction.

The Final Business Case stated that a proper risk management had been followed under the appointment of a Risk Manager (§11.3) and that these procedures were deemed sound by Audit Scotland (§11.6).

The final business case outlines the evolution of the cost estimate for Phase 1a as follows:
- 2005: £484m, incl. risk contingency and optimism bias adjustment of 24%;
- 2006: £500m, incl. 12% risk contingency at P90, but no optimism bias uplift; and
- 2007: £498.1m, incl. 15% risk contingency at P90.

Financial analysis from April 2008 (CEC01425552) shows an anticipated final cost for the project of £508m including a cost risk of £32m at P80, rather than the P90 which the project had used previously. By the close of contract negotiations, these figures had been refined to a budget of £512m including a risk contingency of £31.2m (CEC01338847).

The final business case stated that TIE had established in consultation with Transport Scotland that the 2006 cost estimate did not require an allowance for optimism bias due to adopting a higher than recommended level of confidence in the quantitative risk analysis (§10.14).

The risk contingency of 15% was estimated, using Monte Carlo simulation, to be a P90 estimate. No further adjustment was made for optimism bias. The Final Business Case argued that "Instead of using OB [optimism bias], TS [Transport Scotland] and CEC [The City of Edinburgh Council] adopted a very high confidence figure of 90% (P90) in the estimate of risk allowances to cover for specified, unspecified risk and OB" (§11.42).

The final business case further stated that the total budget of £498.1m included contingency of £49.1m and base estimate of £449.1m (§10.47). The experts noted, that this is inconsistent with the reported figure in the same paragraph that the contingency is 15%. The 15% contingency in the final business case was calculated based only on the construction period cost, and not the total estimated cost.

It should be noted that the STAG 2 appraisal report, prepared in December 2006 together with the first version of the Final Business Case, includes a risk and optimism bias uplift of 16% for the combined Phase 1a and 1b estimate (§7.108 of the STAG 2 report). The STAG 2 report also references the Outline Business Case, which included a comparison of the Edinburgh project to comparable projects. In this comparison, it was noted that:

- Cost overruns were up 25% of award construction cost;
- Projects were typically delayed by three to six months, optimism bias guidance for schedule suggests a 2% uplift, i.e. 1 month on a 39-months construction programme; and
- Cost escalation in utilities diversion have been recognised and are being addressed by including MUDFA in the design process (§10.33 of the STAG 2 report).

The STAG 2 appraisal argues that for the optimism bias uplift the project at OBC stage was classified as a "standard civil engineering" project, which implies a starting value for optimism bias upper bound of 44% and lower bound of 3% (§10.43).

The STAG 2 appraisal further argues that the Mott MacDonald report demonstrated that with effective risk management the optimism bias could reduce to 3%. The STAG 2 appraisal further states that "However, the project's enhanced procurement strategy, which was specifically developed with the consideration of risk, means that it is expected that optimism bias will be near 0% at Contract Award and will come within the 90% confidence level for risk." (§10.45)

At OBC stage TIE estimated that the extensive development work had reduced the optimism bias to 24%, which included a risk allowance for specified risks of circa 10% (§10.46).

The STAG 2 appraisal also noted that concurrent work on the final business case, led to a cost increase, which reduced the risk allowance from 16% to 12%.

The Final Business Case did not include further results of the quantitative risk analysis, other than the stated value of P90 risk. In comparison, the revenue plan in the STAG 2 appraisal contains the full Monte Carlo results of different scenarios.

The 2007 risk review (CEC01496784) summarised the Final Business Case's risk analysis and found that risks remained with the public sector. The review endorsed the £498m risk-adjusted cost estimate at P90 and added a P90 schedule risk estimate of 21 days, which under the contracts was valued at £2.2m.

In addition, the response to the Final Business Case by Transport Scotland (TRS00004270) stated "Transport Scotland perceives 12% risk allowance for a rail-related project to be optimistic, although questioned whether some of this may be included in the base cost." CEC, TEL, and TIE responded: "The process for risk management is defined in the Project Risk Management Plan and related project control procedures as previously shared with Transport Scotland. The risk allowance equating 12% of project base cost represents quantification of the identified risk profile at the time of DFBC [Draft Final Business Case] preparation. The adopted procurement approach resulted in a different risk profile to that of a traditionally procured rail project…" (item 9.1).

# 8. Expert Evaluation

## Treatment of Optimism Bias in the Draft Interim Outline Business Case

The Draft Interim Outline Business Case (30 May 2005) stated that the project sought confirmation from the authors of the Mott MacDonald report that the project should be classified as standard civil engineering rather than non-standard civil engineering project. The business case argues that the data for the non-standard civil engineering projects does not reflect the 40-year experience of building light rail projects in the UK (§6.4.3).

Consequently, the planners expected the project to be a standard civil engineering project, which in the Mott MacDonald study consisted of only three road projects (A34 Newbury bypass, A564 Derby southern bypass, M60 Denton to River Medlock contract 1).

In the experts' view, two points are noteworthy: (1) the planners did not consider all available information by only including the Mott MacDonald data and not the data in Flyvbjerg and COWI (2004), which includes rail projects. (2) Planners implicitly or explicitly assumed that the project's risk is similar to road and not rail projects, which is an unusual assumption. This casts doubt about whether the true risks of the project were understood.

The Draft Interim Outline Business Case argued that optimism bias has reduced in the project (§6.4.3) from 44%, which is the number recommended in the DfT TAG, to 24%. The business case argued that the reduction was measured by evaluating the list of factors contributing to optimism bias, which are listed in the HMT Green Book. The business case points out that this reduction is not due to mitigating individual risks, but due to "progress to varying degrees in the management of all of the 237 identified project risks." (§6.4.3, p. 91).

The planners argued that enhancements to the risk management regime, rather than the mitigation of risks, led them to reduce the optimism bias uplift. However, at the point of the Draft Interim Outline Business Case, the effective risk mitigation was an unproven assumption. While risks were being addressed and the risk regime in general was enhanced, according to the OGC's assessment, whether or not the risks were effectively mitigated, or would be mitigated in the future, could only be proven during construction.

Further, the Draft Interim Outline Business Case reports the results of the quantitative risk analysis of the risk register. At P95 the cost risk was estimated to be 22%. Again, as argued in the analysis of the final business case, this is an indication that the quantitative risk analysis underestimated the tail risk (i.e. risk at levels of P60 and above), and planners should have been aware of data that documented the tail risk.

Arup reviewed the business case preparation (CEC01799560) before it reached outline business case stage. Arup raised the issue that the project reduced the optimism bias uplift. Arup noted that "The project's averaging of mitigation factors is likely to have led to underestimating OB uplifts. Further justification of the likely cost of mitigation strategies should be provided" (§9.18). The Outline Business Case gives this justification but did not fully explain why it reduced the 44% uplift to 24%.

Lastly, the Draft Interim Outline Business Case argued that the headroom between the then base cost estimate and the funding envelope was 54% and thus covered the project's cost plus 44% of optimism bias uplift suggested by the Mott MacDonald data. It should be noted that the DfT guidance on optimism bias procedures available at the time shows that at P80 rail projects ought to use an uplift of 57% (68% at P90, and 80% at P95). At P80 the headroom between the base cost and the funding envelope would have been used up.

In the view of the experts, the Draft Interim Outline Business Case overstates its case with regards to cost risk. The project team argued that it would deliver according to the budget envelope with more than 95% certainty while the data in DfT's Guidance on Optimism Bias, which were available to the planners, suggested that a 20% risk of exceeding the funding envelope existed, i.e. a risk four times higher.

# Treatment of Optimism Bias and Risk in the Final Business Case

At the time of the creation of the Final Business Case (December 2007) the official guidance documented the best available data and knowledge with regards of managing and estimating project risk.

The official guidance recommends steps to implement a good risk management regime. Thus, the various guidance documents asked the right questions to check that the regime is in place, e.g. defined risk management plan, risk management processes, use of risk registers, and use of quantitative risk analysis.

The official guidance documents also require that the risk management regime be checked by independent experts, and that its outcomes are independently assessed. Both steps should in theory provide a project with a robust and working approach to risk management.

Specifically, optimism bias adjustments were conceived to evaluate the questions of economic viability and affordability of a project. In practice, the quantitative risk analysis plus an adjustment for optimism bias is commonly used to set contingencies and plan budgets.

For the Final Business Case, the assessment of risks and optimism bias was important to determine the questions of affordability and viability of the project. The affordability question is evidenced by the analysis in the Final Business Case that the project's headroom was 29% between the base cost estimate and the approved funding envelope. The viability question is evidenced by a benefit cost ratio of 1.77.

In hindsight, it is evident that the risk analysis was insufficient. The Final Business Case document shared the observation that in a small sample of comparative projects[1] no cost overrun greater than 25% was observed.

---

[1] Seven light-rail projects in Dublin, Nottingham, Manchester, Sheffield, Midland Metro, Docklands Light Railway, and the Croydon Tramlink.

The project decided to plan for a risk exposure at P90 instead of the guidance suggested value of mean+6%. The Final Business Case and the STAG 2 appraisal argued that the P90 value is larger than the mean+6% value. In both cases this judgement rests on the quality of the quantitative risk analysis.

To assure the quality of the quantitative risk analysis the project relied on reviews by experts and Transport Scotland. Independent review is, in the experts' opinion, an appropriate step to ensure that the analysis is not underestimating the project's risk exposure. However, since experts are also optimistic, independent reviews are not a guaranteed way to get risks right.

In the experts' view, there are some doubts about the quality of the quantitative risk analysis which estimated a cost risk of 15% at P90, which seems low given the high level of confidence and the evidence in the official guidance documents available to the project at the time, when the project created the Final Business Case. This indicates to the experts that optimism bias is likely to have entered the risk assessment process during quantitative risk analysis, which is understandable, because quantitative risk analysis is based on expert judgement and such judgement has been proven to be prone to optimism bias. Instead of reducing optimism bias, quantitative risk analysis seems to have increased this for the Edinburgh Tram. In addition, the optimism was perpetuated by portraying the cost estimate with a high degree of confidence.

When the project made a numerical argument (that the guidance recommendation of mean QRA + 6% OB uplift is less than the project's P90) to support its approach to risk quantification, in effect the project switched back into a full inside view of the project risks. This, in the view of the experts, is problematic.

Even if the risk analysis calculated the risks accurately, the Edinburgh tram project might have fallen into the top 10% of the estimated risk exposure. Thus, even a high level of confidence is no guarantee of a certain project budget.

However, in the view of the experts, one point of contention is that the project classified itself as a standard civils project in earlier estimates. Light rail schemes are classified in the guidance document as non-standard civils projects. However, the resulting difference is small, i.e. P(Mean) + 6% versus P(Mean) + 3% (DfT TAG, p. 27, table 9). In the view of the experts it is more problematic that the expected reduction in optimism bias is based only on a small data set, i.e. the Mott MacDonald report, and the assumption that optimism can be reduced through risk mitigation measures.

At the time when the Final Business Case was produced, the risk management system established by the Edinburgh Tram project addresses the requirements of the official guidance documents, i.e. defined processes, use of a risk register, and a risk management plan. The details provided describe a risk management system that is conventional and typical for projects of this kind, in the view of the experts.

STAG and TAG guidance in place at the time when the Final Business Case was produced required a quantitative risk analysis. The process described in the Final Business Case documents (i.e. use of risk registers, valuation of risks through the impact multiplied by likelihood approach, use of Monte Carlo) followed the recommendations made in the official guidance, in the view of the experts.

However, the experts note that the quality of the quantitative risk assessment seems problematic. The produced estimate of a cost risk exposure of 15/12% of the base cost and schedule risk exposure of 2% (21 days) both at a 90% level of confidence (P90) seems unreasonably low. P90 means it is assumed there is only a 10% chance that each of these estimates will be exceeded.

The P90 adopted by the project is an unusually high level of confidence; the DfT guidance suggest P80 as the conservative value.

In the experts' experience, Monte Carlo-based quantitative risk analyses are only as good as the data they are based on, and such data commonly underestimate the tail of the risk exposure. Based on empirical data, the DfT Guidance on optimism bias shows that P90 in rail projects approximately equates a cost risk exposure of 70%. This shows a clear gap that the project should have been aware of between its own analysis of the tail risk and the data that were available.

It should be noted, that the figures presented in the DfT Guidance are based on the approved business cases of these projects (i.e. final business case at the time of decision to build). In contrast, the TAG guidance applies these data to the outline business case (GRIP stage 3) and at final business case stage (GRIP stage 5). It expects a reduction in cost risk exposure to a level of the mean estimate of the quantitative risk analysis + 6%. Those figures are based on the findings of the Mott MacDonald study.

The guidance suggests that "mean QRA + 6% uplift on the risk-adjusted scheme cost" should result in a reduction of the risk exposure compared to the 40% uplift at the previous stage of project development. However, no guidance is given as to what range of risk exposure would be expected after adjusting the quantitative risk analysis in this way. This is likely to contain an element of optimism bias in its own right, in the experts' judgement.

With regards to the optimism bias adjustment, the experts noted that the project was aware of the requirements made by the official guidance: earlier project cost risk estimates (outline business case, initial business case) included an optimism bias uplift. In the Final Business Case the project decided to adopt a different approach. Instead of uplifting the mean of the Monte Carlo analysis by 6%, the project chose to estimate risk at P90. The STAG 2 appraisal argues that the P90 figure is greater than the figure that would have resulted from following the guidance (mean+6%). Only limited results of the quantitative risk analysis are included in the Final Business Case documentation.

The QRA dated 08 December 2007 (CEC01397542) documents that the mean risk estimate for phases 1a and b was £38.6m and the P90 risk estimate was £51.6m. The final business case estimated the total cost of phases 1a and 1b at £580m (Final Business Case §1.65). The experts infer from this that the base cost estimate was £528.4m (£580m-£51.6m). The QRA states that the mean risk estimate was £38.6m. Thus, the risk-adjusted total cost of phases 1a and 1b add up to £567.0m (base cost + mean risk estimate). The 6% uplift applied to the risk-adjusted scheme cost equals £34.0m (6% of £567.0m). Thus, the total risk provisions of mean QRA and the OB uplift should have been £72.6m (£38.6m + £34.0m).

|  | GBP mio | Source |
|---|---|---|
| Total cost | 580 | Final business case §1.65 |
| - P90 risk estimate | 51.6 | QRA 08-Dec-07 |
| = Base cost | 528.4 |  |
|  |  |  |
| Base cost | 528.4 | from above |
| + Mean risk estimate | 38.6 | QRA 08-Dec-07 |
| = Risk-adjusted scheme cost | 567.0 |  |
|  |  |  |
| 6% of the risk-adjusted scheme cost | 34.0 |  |
| + Mean risk estimate | 38.6 | QRA 08-Dec-07 |
| = Risk estimate mean + 6% uplift | 72.6[2] |  |

The P90 risk estimate in the December 2007 QRA is £51.6m. The experts find that the P90 estimate is not larger than the mean + 6% figure (£72.6m). While the experts lack the risk estimates to ascertain whether this was true for the STAG 2 appraisal, where the argument was made by the project, at least in the final business case the argument for abandoning OB considerations in favour of solely relying on QRA is inaccurate.

Moreover, the project's justification for this seemingly low estimate of the cost risk exposure was centred around two points; (1) that utility diversion (MUDFA) works had already commenced; and (2) that the procurement strategy would significantly de-risk the project. The project expressed the view that at contract award, optimism bias is 0%. In other words, the project assumed that due to ongoing refinements of its risk register and its risk analysis and due to its commercial strategy, all risks will be fully known. This indicates to the experts that the risk management team did not fully understand the nature of optimism bias and, because of this, the team and project would have been particularly prone to such bias.

It should be noted that the Mott MacDonald review, which the project will have been aware of, questions this assessment. Firstly, Mott MacDonald clearly states that their analysis estimated that procurement accounts for 2% of the total optimism bias uplift, which indicates that the project should have been aware of evidence that procurement strategies might have less of an influence on decreasing risks than argued in the Final Business Case. Secondly, Mott MacDonald's analysis also suggests that at contract award the optimism bias falls to the lower bound of the estimate (i.e. 6%). In other words, that unknown risks account for 6% of the risk-adjusted cost estimate.

---

[2] The base cost estimate is not stated in the QRA document, hence we applied the calculation to the more conservative figure of £580m for parallel construction of phases 1a and 1b. If the slightly higher £585m estimated cost for sequential construction of phases 1a and 1b are used, then the mean risk + 6% uplift adds up to £72.9m. We noted that the December 2007 estimate considered only phase 1a. In the risk estimates the planners calculated the cost risk by allocating different shares of each risk to both phases. Thus in principle the above finding for phases 1a and 1b also holds true for the consideration of phase 1a only.

Moreover, the Final Business Case indicates that the project was aware of the interface risks between the three key contracts (design, utilities, and infrastructure construction). The Final Business Case clearly states the retained risks as a result of the suggested commercial structure, including the novation option to remove the interface between the design and infrastructure construction contract.

## Summary Evaluation

In the experts' judgement, the approach taken to estimates, risk and optimism bias in the Edinburgh tram project was generally similar to the approach of other projects of a similar nature. Equally, the mitigation measures planned and the work to understand risk were similar to those of other projects.

However, the project relied solely on an inside view to assess risk. The OGC reviews and the Monte Carlo simulation confirmed the assessment but did not remove optimism bias from the estimates.

Furthermore, the risks identified in the contracting strategy in the final business case combined with the early-warning signs of delays in the design work and the utilities diversion project ought to have led to a more cautious approach regarding risks and uplifts for the overall project.

# 9. Recommendations

## Recommendations to Improve Official Guidance

In the view of the experts, the guidance is appropriate for major projects today and can help a project to establish a de-biased estimate of risk. However, the history of the Edinburgh tram project points to necessary enhancements and potential pitfalls in applying the guidance.

First, the official guidance works on the assumption that a high-quality risk analysis has taken place to appraise projects at outline business case and final business case stage. The guidance requires supporting evidence and expert reviews to assure the quality of risk analyses. However, the history of the Edinburgh tram project demonstrates three points:

1. The Edinburgh project established a risk management regime (systems, tools, processes) that was in line with typical risk management regimes of UK infrastructure projects at the time. The Edinburgh tram project shows that conventional risk management, despite the best intentions, is not getting risks right;
2. Specifically, quantitative risk analysis following the inside view (i.e. risk register plus Monte Carlo simulation) is insufficient to create extreme downside scenarios (i.e. P90 or above) in order to appraise a project's viability and affordability, and, by creating a false sense of certainty, may add risk instead of reducing it, as appears to have been the case in Edinburgh; and
3. Optimism in expert reviewers is difficult to root out, unless all analyses are based on hard, empirical data, i.e. taking out human judgement as much as possible.

Second, DfT's guidance at the writing of this report (TAG Unit A5.3, December 2015) is based on the same data as the 2006/2007 guidance (Flyvbjerg and COWI 2004, TAG, and STAG). To decide whether the guidance is appropriate to major projects today, one would need to evaluate whether the data are still relevant, which was done as part of preparing this report.

The comparison of the data available today for rail projects and the data provided by Flyvbjerg and COWI (2004), which is used in the most recent DfT guidance (TAG Unit A5.3, December 2015) shows a different risk profile. The P40 is lower and the P80 higher in the updated reference class than in the 2004 reference class for rail projects. The 2016 data set must be considered more reliable than the 2004 data set, because it is significantly larger and more up to date.

The Edinburgh tram project illustrates that the guidance needs to be improved on how to combine optimism bias and QRA. In particular, the 6% uplift on the risk-adjusted scheme cost at the mean risk estimate produces figures that too low and not supported by the data.

These finding support four recommendations:
1. The data and recommended uplifts in the DfT's TAG might need to be updated. More recent and larger data sets are available;
2. The guidance should be improved on how optimism bias uplifts should be combined with QRAs;

3. Projects could establish their own, more precise, reference class. Sufficient data is available to construct reference classes that focus on specific project types;
4. However, if projects construct their own reference class, statistical analysis should be used to decide which project types to include in the reference class. For example, the light rail projects above are statistically similar to other rail projects and therefore a reference class only of light rail projects would make the error of discarding valuable information.

Third, the analyses behind the TAG guidance measured actual outcome data against different baselines. The baseline in the two data sources and in TAG are not consistent: Mott MacDonald (2002), Flyvbjerg and COWI (2004) measure cost overruns based on the final decision to build (i.e. the final business case). TAG uses those numbers as uplifts for the outline business case stage. Full distributional information for the different baselines is needed.

# Recommendations for Project Funders, Sponsors and Project Managers

Project cost estimates ought to be adequately challenged and controlled. The key questions funders and sponsors need to consider are whether a project's base estimates are robust and whether the risk estimates are biased.

Funders, sponsors and project managers can challenge and evaluate the quality of estimates by taking the outside view and comparing the outside with the inside view.

Funders and project sponsors need to understand their risk appetite in order to evaluate the three key questions of project appraisal:
- Is the project economically viable?
- Is the project affordable?
- What project budget and what contingencies should be allocated to different levels, i.e. builders, project director, owner?

The risk appetite and hence the total cost estimate will differ for each of these questions. Sponsors and funders should use probabilistic forecasts instead of single point forecasts to capture this reality.

In general, project funders, sponsors and project managers should be cautious when adjusting uplifts. In addition, project risks are only removed when the project is realised. The Edinburgh tram project shows that a commercial strategy and contractual risk transfer do not necessarily eliminate risk.

The Edinburgh tram project also shows that risk management and other functions need to be joined up. Establishing a clear line of sight from the strategic benefits down to requirements, design, and delivery and commercial strategy. The Risk Management Module of the Project Initiation Routemap (IPA 2016) provides guidance to build a joined-up risk management system.

Projects with very large cost overruns are also called Black Swans; i.e., extreme events with massively negative and unforeseeable outcomes. Managers tend to ignore this. They are treating projects as if they exist largely in a deterministic Newtonian world of narrow variations between cause and effect. Project funders, sponsors and project managers should pay careful attention to their tail risk, i.e. the domain of Black Swans. Early warning indicators, active attempts to break down project size, removing project complexity, and preparing project responses to and recovery from adverse events can reduce the impact of extreme events.

During delivery, effective governance needs to provide constant challenge and control of the project, including recording of where the project is compared to its baseline, while at the same time enabling problem solving, including quickly getting the project back on track, whenever it begins to veer off course.

To provide adequate challenge and control, the governance bodies need to receive unbiased and up-to-date information about project performance. In similar projects (Flyvbjerg and Kao 2014) the experts found that effective governance relies on multiple channels of information to senior decision makers, for example, data-driven reports on project performance and forecasts combined with reports by the management team and independent audits. In the reporting, special emphasis must be placed on detecting early-warning signs that cost, schedule and benefit risks may be materialising, as they tend to do, so damage to the project can be prevented. When early-warning signs emerge, projects should revisit their assumptions and reassess risk and optimism bias forecasts.

Project sponsors and funders should critically review claims mitigation measures have reduced project risk, especially when they are based on assumptions and not proven. The Edinburgh Tram claimed to mitigate risk through an un-tested contract approach. Generally, first-time innovation should be seen as adding, not removing, risk.

A key practical challenge is that top-management governance bodies often include representatives without prior experience in managing major projects. Thus, effective communication between the project and its governance bodies becomes a challenge. Closing this capability gap but also carefully designing management information is important (Flyvbjerg and Kao 2013).

Effective governance constantly challenges and controls and is thus a source of conflict. However, research shows that accountability and creating a safe space to raise difficult issues go hand in hand (Edmondson 2003). Thus, governance bodies need to clearly acknowledge constraints, frame issues accurately, embrace reporting of issues and risks, positively engage with messengers, and encourage dissent and the communication of bad news.

The Edinburgh Tram project shows that the key issue of providing independent oversight and external validation is the quality of evidence. The Edinburgh Tram was reviewed by outside parties (Office of Government Commerce, KPMG etc.) yet risks might have been overlooked.

The project showed that not the process but the quality of evidence is key to providing effective oversight and validation. Garbage in, garbage out, here as elsewhere. Optimism bias uplifts are aimed at providing quality evidence by eliminating psychological and political bias in project plans.

Incentive structures are a root cause of the typical challenges of cost overruns, delays and benefit shortfalls faced by projects. Project promoters are strongly incentivised to portray a proposal as positively as possible to get project approval.

This conflict of interest can be addressed by using third party or external validators of project proposals. The NAO (2013) review of evaluation activities across government found that a variety of constructs are being used ranging from fully internal reviews to fully external ones. The key challenge is to ensure autonomy, and thus freedom of political biases, with the ability to access commercially sensitive information and providing a high-quality evaluation.

For example, the UK Ministry of Defence – a department with a large portfolio of major projects – has set up an independent cost accounting and assurance service, which subjects all project cost estimates to an independent assessment. Some projects, for example Heathrow Terminal 5 as discussed above, have appointed independent cost estimators as part of the project organisation.

In the experts' view, while value-for-money is a constant concern in public projects, ensuring high-quality evidence and independent oversight are the key problems that must be solved in order to prevent costly break-fix projects that destroy value instead of creating it.

# Appendix I – List of Reviewed Documents

| Begin Prod ID | Document Description | Sort Date |
|---|---|---|
| CEC00380894 | TIE: Edinburgh Tram Progress Report, September 2005 | 30/09/2005 |
| CEC00380898 | Edinburgh Tram Network: Outline Business Case, Draft for Discussion, March 2006 | 30/03/2006 |
| CEC00455240 | Edinburgh Tram Project - Tram Lines 1 and 2 Meeting of the Council, 11 December 2003 | 11/12/2003 |
| CEC00630633 | Preliminary Financial Case – Update Line One | 01/09/2004 |
| CEC00632759 | Edinburgh Tram Network: STAG Appraisal: Line One | 28/11/2003 |
| CEC00640848 | Edinburgh Tram Network: STAG 2 Appraisal Report, prepared by Steer Davies Gleave, Dec 2006 | 30/12/2006 |
| CEC00642726 | Edinburgh Tram Network: STAG 2 Appendices Line One | 28/11/2003 |
| CEC00642799 | Edinburgh Tram Network: Preliminary Financial Case:Update: Line Two | 01/09/2004 |
| CEC00643516 | Edinburgh Tram Network: Final Business Case v 2, 7th December 2007 | 07/12/2007 |
| CEC00906940 | CEC: Edinburgh Tram : Financial Close and Notification of Contract Award | 01/05/2008 |
| CEC01019126 | Arup Scotland: Edinburgh Tram Line 2 Review of Business Case: Final | 26/10/2004 |
| CEC01190799 | Arup Transport Planning: Edinburgh LRT Masterplan Feasibility Study: Final Report (Jan 2003) | 29/01/2003 |
| CEC01395434 | Edinburgh Tram Network: Final Business Case v2, 7th December 2007 | 07/12/2007 |
| CEC01496784 | TIE: Project Risk Review Report: Readiness Review; Oct 2007 | 14/10/2007 |
| CEC01562064 | TIE Project Gateway 3 Review: Readiness Review; Oct 2007 | 09/10/2007 |
| CEC01623145 | TIE: Integrated Transport Initiative for Edinburgh and South East Scotland: A Vision for Edinburgh | 30/09/2002 |
| CEC01629382 | TIE Project Readiness Review: Readiness Review issued to Transport Scotland: September 2006 | 28/09/2006 |
| CEC01649235 | Edinburgh Tram Network: Final Business Case version 1, 3 October 2007 | 03/10/2007 |
| CEC01791014 | TIE Project Gateway 2 Review: Follow Up Report: Issued to Transport Scotland; 22 November 2006 | 22/11/2006 |
| CEC01793454 | TIE Project Readiness Review: Issued to Chief Executive: 25 May 2006 | 25/05/2006 |
| CEC01799560 | Arup Scotland: Scottish Parliament: Edinburgh Line 1: Review of Final Business Case, Oct 2004 | 26/10/2004 |
| CEC01821403 | Edinburgh Tram Network: Draft Final Business Case; November 2006 | 30/11/2006 |
| CEC01836749 | Edinburgh Tram Network: STAG 2 Report: Line Two; September 2004 | 01/09/2004 |
| CEC01868789 | City of Edinburgh Council: Strategic Project Review, | 01/09/2002 |

| | prepared by Turner & Townsend, September 2002 | |
|---|---|---|
| CEC01875336 | Edinburgh Tram Network: Interim Outline Business Case, Draft for Discussion; 30 May 2005 | 30/05/2005 |
| CEC01916700 | Waterfront Edinburgh: Feasibility Study for a North Edinburgh Rapid Transit Solution ; July 2001 | 01/07/2001 |
| CEC02044271 | CEC: Edinburgh Tram Project: report no CEC/22/11-12/CD; 30 June 2011 | 30/06/2011 |
| CEC02083184 | CEC: Edinburgh Tram Project – Update Report; 24 June 2010 | 24/06/2010 |
| CEC02083448 | CEC: Edinburgh Trams Contracts Acceptance; Report 139; 20 December 2007 | 20/12/2007 |
| CEC02083466 | CEC: Edinburgh Tram Draft Final Business Case; 21 December 2006 | 21/12/2006 |
| CEC02083538 | CEC: Edinburgh Tram Final Business Case; 25 October 2007 | 25/10/2007 |
| CEC02083547 | CEC: Edinburgh Tram; Report; 26 January 2006 | 26/01/2006 |
| CEC02084255 | Department for Transport: Transport Analysis Guidance (TAG): Unit 3.5.9 The Estimation and Treatment of Scheme Costs; September 2006 | |
| CEC02084256 | HM Treasury: The Green Book: Appraisal and Evaluation in Central Government; 2003 edition. | |
| CEC02084257 | The British Department for Transport: Procedures for Dealing with Optimism Bias in Transport Planning: Guidance Document: June 2004 | 10/06/2004 |
| CEC02084489 | STAG: Scottish Transport Appraisal Guidance: Chapter 12 Risk and Uncertainty | 01/09/2003 |
| CEC02084689 | Mott MacDonald: Review of Large Public Procurement in the UK, July 2002 | 25/04/2002 |
| CEC01300167 | Spreadsheet: Phase 1A – Budget at Financial Close | |
| DLA00004903 | TIE: Prequalification Guide for Edinburgh Tram Network: Guide for candidates seeking to bid for the development, partnering and operating franchise agreement (DPOFA) | 08/06/2003 |
| TRS00000016 | TIE: Edinburgh Tram Network - Preliminary Financial Case_ Line Two; 4th December 2003 | 04/12/2003 |
| TRS00000041 | TIE: Edinburgh Tram Network - STAG Appraisal- Line One; 30th July 2004 | 30/07/2004 |
| TRS00000054 | TIE: Edinburgh Tram Network: Preliminary Financial Case: Line One; 4th December 2004 | 04/12/2003 |
| TRS00004270 | Edinburgh Tram Network: Draft Final Business Case: Combined response from CEC, TEL and TIE to T.S. | |
| TRS00011725 | CEC: Edinburgh Tram Project: Report No CEC/39/11-12/CD; dated 25 August 2011 | 25/08/2011 |
| TRS00018617 | TIE: Edinburgh Tram Network: STAG Report: Line Two; 31st March 2004 | 31/03/2004 |

| CEC01300167 | Financial and risk summary | 01/03/2008 |
| CEC01425552 | Financial and risk summary [same as CEC01300167] | 01/03/2008 |
| CEC01338847 | Edinburgh Tram project - financial close process and record of recent events | 12/05/2008 |
| CEC01397542 | Edinburgh Tram project – risk allocation report | 26/11/2007 |
| CEC01397541 | Risk exposure graph phase 1A P90 | 26/11/2007 |

# Appendix II – Scope and Background of the Report

## The Edinburgh Tram Project

The Edinburgh Tram Project was set up to plan and construct a tramway in Edinburgh, Scotland.

The project was first proposed in June 2000, construction started in 2008, and the tram line opened on 31 May 2014.

The tram services the 14 km between York Place in New Town and Edinburgh Airport, including 15 stops in total.

In 2007, in the accepted version of the final business case, the cost of the tram was estimated at £498.1 million (in nominal cost) and the planned opening date was estimated as Q1 2011. The tram opened 3 years late with a final outturn cost of £776 million (nominal cost).

*Figure 1 Alignment of the Project*

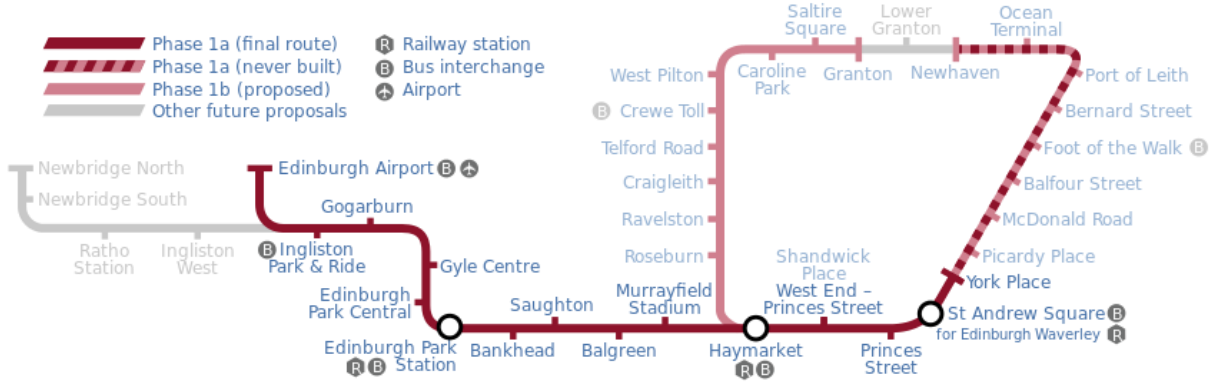

Source: The City of Edinburgh Council, 2013.

## Timeline of the Project

The following timeline of the project outlines key steps of the project planning phase and gives an overview of project execution with regards to the documents that were analysed for this report. It is important to note that throughout the project, estimates were produced in nominal terms, i.e. including effects of inflation.

| | |
|---|---|
| June 2000 | CEC publishes its Local Transport Strategy which sets out that the development of a tram network is central to its transport policy. |
| October 2000 | CEC approves the Local Transport Strategy. |
| April 2001 | CEC commission feasibility studies into an Edinburgh tram system. |

| July 2001 | Waterfront Edinburgh Ltd produces the Feasibility study. |
| --- | --- |
| May 2002 | CEC establishes TIE as an arms-length company to investigate how best to deliver its local transport strategy. |
| September 2002 | TIE submits its proposals to CEC. Turner and Townsend produces a Strategic Project Review for CEC. |
| January 2003 | Arup Transport Planning produces the **Final Feasibility Study**, incl. a total cost estimate of £465.55 million. |
| March 2003 | Scottish Ministers announce £375 million available in principle for tram system. |
| November 2003 | **STAG appraisal for line 1** is published. |
| December 2003 | STAG appraisal is sought for lines 1 and 2. Preliminary financial cases are published for lines 1 and 2. |
| January 2004 | Two Bills submitted to the Scottish Parliament intended to enable the construction of the tram system. |
| March 2004 | **STAG appraisal for line 2** is published. |
| July 2004 | Further STAG appraisal for line 1 is conducted. |
| September 2004 | Update of the Preliminary Financial Case for lines 1 and 2. Updates of STAG appraisal for line 2. |
| October 2004 | Ove Arup and Partner Ltd produce a business case review for lines 1 and 2. |
| May 2005 | **Draft Interim Outline Business Case** |
| September 2005 | TIE appoints Parsons Brinkerhoff to facilitate the early identification of utility diversion works and completion of design drawings. TIE provides a progress report for the Scottish Parliament. |
| January 2006 | Report to the Council makes recommendations for funding and phasing. |
| March 2006 | **Draft Outline Business Case** is produced. Bills receive Royal Assent. |
| October 2006 | TIE appoints Alfred McAlpine Infrastructure Services to be responsible for the diversion and protection of utilities along the tram route. |
| December 2006 | **STAG 2 appraisal** is conducted. **Draft Final Business Case** is published. Joint report to the CEC seeks approval of the Final Business Case. |
| June 2007 | Auditor General publishes their report 'Edinburgh transport projects review' which includes the tram project. Following a debate and vote, the Scottish Parliament calls on the SNP administration to proceed with the Edinburgh tram project within the budget limit set by the previous administration. |
| October 2007 | **Final Business Case**, version 1 is published. TIE signs pre-contract agreements for the supply and maintenance of 27 tram vehicles with Construcciones y Auxiliar de Ferrocarriles SA (CAF). TIE announces the consortium Bilfinger Berger Siemens (BBS) as the preferred bidder for construction of the tram infrastructure, including rails, overhead power cables and a tram depot. |
| December 2007 | **Final Business Case**, version 2 is published. TIE signs a mobilisation and advance work agreement for infrastructure construction with BBS. CEC approves the final business case. |

| January 2008 | Scottish Ministers offer grant support for Phase 1a. |
|---|---|
| May 2008 | Tom Aitchison produces report to the CEC.<br>BBS appointed as contractor for the construction of the tram infrastructure. |
| February 2009 | Major dispute arises between BBS and TIE, one week before track-laying work was due to start in Princess Street, amid claims that BBS is seeking an additional £50-80 million funding. |
| April 2009 | CEC announces that, in view of the economic downturn, Phase 1b of the project is not proceeding in the foreseeable future. |
| June 2009 | A week of informal mediation is held between TIE and BBS, which examines, among other things, the interpretation of key clauses in the pricing schedule, risk allocation and the substantiation of changes and value engineering issues. |
| July 2009 | TIE reports to the Tram Project Board (TPB) that the mediation had not been successful. TPB endorses TIE's strategy of adopting a more formal approach to managing the contract. |
| November 2009 | Carillion (owner of Alfred McAlpine since December 2007) completes its works package of diverting 40,000 metres of utility pipes and cables. TIE appoints Clancy Docwra and Farrans to divert the remaining 10,000 metres. |
| December 2009 | Following further disputes with BBS, the TPB concurs with TIE's proposal that, in view of lack of progress, a fundamental review of the contractual position with BBS should be conducted. If required, formal legal processes should be started to bring the major issues to a head to allow the project to progress. |
| March 2010 | TIE informs CEC who tells Transport Scotland that it is unlikely that all of Phase 1a of the project can be delivered for £545 million. £348 million has been spent on the project up to that point. |
| March 2010 | The TPB approves TIE's strategy for the future direction of the project including management of the infrastructure construction contract with BBS. |
| June 2010 | Directors of City Development and Finance report on the project.<br>CEC reports to full council meeting on progress of the project. Council requests a refreshed business case detailing the capital and revenue implications of all options being investigated by TIE. |
| October 2010 | CEC provides an update on progress and outlines an incremental approach to the project which would see the opening of a line from Edinburgh Airport to St Andrew Square as the first phase. No cost or benefit figures are provided and the council requests a further report to be prepared for its December 2010 meeting. |
| October 2010 | The Accounts Commission and the Auditor General for Scotland announce their intention to carry out a further review which will provide an independent commentary on the Edinburgh tram project's progress and costs to date and its governance arrangements. |
| December 2010 | A refreshed tram business case is presented. The report includes considerations of incremental delivery of Phase 1a, an update on the economic case for Phase 1a, expenditure to date and an assessment of funding and affordability. The council also notes that a report would be |
|   | submitted within one year on the operational and governance arrangements necessary to secure the integration of bus and tram services. |
| June 2011 | Director of City Development reports to CEC with revised plans for the project.<br>Atkins provides an independent review of the Business Case.<br>Revised plans are approved, shortening the line to Edinburgh Airport to St Andrew Square at a revised cost estimate of £770 million. |
| August 2011 | Faithful and Gold provide a validation of the base budget and proposed risk allowance.<br>CEC revises the alignment to run from Edinburgh Airport to Haymarket. This decision was withdrawn 1 week later.<br>TIE is disbanded. |
| November 2011 | CEC further revises the alignment to run from Edinburgh Airport to York Place. |
| November 2011 | Announcement of new opening date in 2014. |
| August 2013 | Testing commences. |
| December 2013 | Full-line testing commences. |
| May 2014 | Opening of the line. |

# Scope of This Report

This report was commissioned by the Edinburgh Tram Inquiry to provide expert input regarding subjects of official guidance, risk and optimism bias, the approach to risk and optimism bias taken by the Edinburgh Tram project, general views on what steps should be taken to ensure effective project management and effective governance of major infrastructure projects.

This report focuses primarily on the Final Business Case, which was first published as a draft in December 2006, then updated in October 2007 and December 2007, when it was approved by CEC.

This report analyses the documents listed in Appendix II. The Edinburgh Tram Inquiry asked the authors of this report to answer specific questions regarding:
- The Key Concepts;
- Probability and risk;
- Assessment of risk;
- Change to OB throughout a project;
- Assessment of OB;
- Assessment of Guidance;
- The means by which tie set its risk and OB allowances in the project budgets;
- Comment on the approach taken by tie to setting its risk and contingency allowances;
- Recommendations to improve existing guidance; and
- Recommendations to funders, project sponsors and project managers.

TRI00000265_0037

# Appendix III – References

# Appendix IV – Experience and Expertise

Bent Flyvbjerg is the first BT Professor and inaugural Chair of Major Programme Management at Oxford University's Saïd Business School and a Professorial Fellow of St Anne's College, Oxford. His main areas of expertise are megaproject management and research methodology.

Flyvbjerg is the most cited scholar in the world in megaproject management, and among the most cited in social science methodology. He is the author or editor of 10 books and more than 200 papers in professional journals and edited volumes. His publications have been translated into 20 languages.

Flyvbjerg serves as advisor and consultant to government and business, including the US and UK governments and several Fortune 500 companies. He is an external advisor to McKinsey and other consultancies. He has worked on some of the largest projects in the world, on all aspects from front-end planning, delivery, and rescue of failing projects.

Flyvbjerg's research has been covered by Science, The Economist, The Financial Times, The Wall Street Journal, The New York Times, China Daily, The BBC, CNN, Charlie Rose, and many other media. He is a frequent commentator in the news.

Flyvbjerg has received numerous honors and awards, including Harvard Business Review's 'Idea Watch' for the most important new idea to follow and the Project Management Institute's and Project Management Journal's 'Paper of the Year Award.' Flyvbjerg was twice a Fulbright Scholar and received a knighthood in 2002.

## CV

**Work History**
- 2012 – Current    Chairman, Oxford Global Projects Ltd.
- 2009 – Current    Professor, Saïd Business School, University of Oxford
- 1993 – 2009       Professor, Aalborg University
- 2006 – 2009       Professor, Delft University
- 1994              Visiting Fellow, European University Institute, Florence
- 1988 – 1992       Senior Associate Professor, Aalborg University
- 1985 – 1988       Distinguished Research Scholar, National Science Council, Copenhagen
- 1985–1986         Post-doctoral Fellow, University of California at Los Angeles
- 1983 – 1988       Associate Professor, Aalborg University
- 1981 – 1982       Doctoral Fellow, University of Aarhus
- 1979 – 1982       Assistant Professor, Aalborg University

**Degrees**
- Dr. Scient., higher doctorate in science, Aalborg University, 2007.
- Dr. Techn., higher doctorate in engineering, Aalborg University, 1991.
- Ph.D. in economic geography, University of Aarhus, University of California at Los Angeles, 1985.

# Publications

"Oxford Handbook of Megaproject Management." Oxford University Press, 2017.

"Big is Fragile: An Attempt at Theorizing Scale." Co-authors: Atif Ansar, Alexander Budzier, and Daniel Lunn. In: The Oxford Handbook of Megaproject Management, Oxford University Press, 2017.

"Does Infrastructure Investment Lead to Economic Growth or Economic Fragility? Evidence from China." Co-authors: Atif Ansar, Alexander Budzier, and Daniel Lunn. Oxford Review of Economic Policy, vol. 32, no. 3, September 2016, pp. 360-390.

"The Oxford Olympics Study 2016: Cost and Cost Overrun at the Games." Co-authors: Allison Stewart and Alexander Budzier. Saïd Business School Workingpaper, University of Oxford, July 2016.

"Reference Class Forecasting for Hong Kong's Major Roadworks projects." Co-authors: Chi-keung Hon, Wing Huen Fok. Proceedings of the Institution of Civil Engineers-Civil Engineering, pre-print publication, June 2016.

"The Fallacy of Beneficial Ignorance: A Test of Hirschman's Hiding Hand." World Development, vol. 84, May 2016, pp. 176-189.

"Tension Points: Learning to Make Social Science Matter." Co-authors: Todd Landman and Sanford F Schram. Critical Policy Studies, 2016.

"The Principle of the Malevolent Hiding Hand; or, the Planning Fallacy Writ Large." Co-author: Cass R. Sunstein. Social Research, September 1, 2015.

"Decision-Making and Major Transport Infrastructure Projects: The Role of Project Ownership." Co-author: Chantal C. Cantarelli. In Hickman, R., Givoni, M., Bonilla, D. & Banister, D. (eds.). Handbook on Transport and Development, Cheltenham: Edward Elgar, pp. 380-393. August 2015.

"More on the Dark Side of Planning: Response to Richard Bolan." Cities, vol. 42, pp. 276-78.

"Why Do Projects Fail?" Co-author: Alexander Budzier, Project Magazine, Summer 2015.

"Report to the Independent Board Committee." Co-authors: Tsung-Chung Kao and Alexander Budzier. MTR Corporation, September 2014.

"Should We Build More Large Dams? The Actual Costs of Hydropower Megaproject Development." Co-authors: Atif Ansar, Alexander Budzier, and Daniel Lunn. Energy Policy no. 69, June 2014, pp. 43-56.

"Overspend? Late? Failure? What the Data Say about IT Project Risk in the Public-Sector." Co-author: Alexander Budzier. In: Commonwealth Secretariat, ed., Commonwealth Governance Handbook 2012/13: Democracy, Development, and Public Administration, London.

"The Project Interview." Project Magazine, April 2013.

"Geographical Variation in Project Cost Performance: the Netherlands versus Worldwide." Co-authors: Chantal C. Cantarelli and Søren L. Buhl. Journal of Transport Geography, September 2012, pp. 324-331.

"Different Cost Performance: Different Determinants? The Case of Cost Overruns in Dutch Transportation Infrastructure Projects." Co-authors: Chantal C. Cantarelli, Bert van Wee and Eric JE Molin. Transport Policy, vol. 22, July 2012, pp. 88-95.

"Characteristics of Cost Overruns for Dutch Transport Infrastructure Projects and the Importance of the Decision to Build and Project Phases." Co-authors: Chantal C. Cantarelli, Bert van Wee and Eric JE Molin. Transport Policy, vol. 22, July 2012, pp. 49-56.

"Olympic Proportions: Cost and Cost Overrun at the Olympics 1960–2012." Co-author: Allison Stewart. Saïd Business School Working Paper, University of Oxford, June 2012.

"Why Mass Media Matter to Planning Research: The Case of Megaprojects." Journal of Planning Education and Research, vol. 32, no. 2, June 2012, pp. 169-181.

"Why Mass Media Matter, and How to Work with Them: Phronesis and Megaprojects." In: Real Social Science: Applied Phronesis. Cambridge University Press, April 2012, pp. 133-171.

"Important Next Steps in Phronetic Social Science." Co-authors: Todd Landman and Sanford F Schram. In: Real Social Science: Applied Phronesis. Cambridge University Press, April 2012, pp. 285-297.

"Introduction: New Directions in Social Science." Co-authors: Todd Landman and Sanford F Schram. In: Real Social Science: Applied Phronesis. Cambridge University Press, April 2012, pp. 1-14.

"Real Social Science: Applied Phronesis." Co-authors: Todd Landman and Sanford F Schram. Cambridge University Press, April 2012, pp. 319.

"Why Your IT Project May Be Riskier Than You Think." Co-author: Alexander Budzier. Harvard Business Review, vol. 89, no. 9, September 2011, pp. 601-603.

"Double Whammy: How ICT Projects are Fooled by Randomness and Screwed by Political Intent." Co-author: Alexander Budzier, Saïd Business School Working Paper, University of Oxford, August 2011.

"Comparison of Capital Costs per Route-Kilometre in Urban Rail." Principal author: Bent Flyvbjerg; co-authors: Nils Bruzelius and Bert van Wee. European Journal of Transport and Infrastructure Research, vol. 8, no. 1, March 2008, pp. 17-30.

"Phronetic Organizational Research." In Richard Thorpe and Robin Holt, eds., The Sage Dictionary of Qualitative Management Research. Los Angeles: Sage Publications, 2008, pp. 153-155.

Hugo Priemus, Bent Flyvbjerg, and Bert van Wee, eds., Decision-Making On Mega-Projects: Cost-Benefit Analysis, Planning, and Innovation.Cheltenham, UK and Northampton, MA, USA: Edward Elgar, 2008, 352 pp.

"Introduction: Scope of the Book." Co-authors Hugo Priemus and Bert van Wee. In Hugo Priemus, Bent Flyvbjerg, and Bert van Wee, eds., Decision-Making On Mega-Projects: Cost-benefit Analysis, Planning, and Innovation. Cheltenham, UK and Northampton, MA, USA: Edward Elgar, 2008, pp. 1-20.

"Public Planning of Mega-projects: Overestimation of Demand and Underestimation of Costs." In Hugo Priemus, Bent Flyvbjerg, and Bert van Wee, eds., Decision-Making On Mega-Projects: Cost-benefit Analysis, Planning, and Innovation. Cheltenham, UK and Northampton, MA, USA: Edward Elgar, 2008, pp. 120-144.

"Aristotle, Foucault, and Progressive Phronesis," in Jean Hillier and Patsy Healey, eds., Critical Essays in Planning Theory, vol. 2, Political Economy, Diversity, and Pragmatism. London: Ashgate, 2008.

"Curbing Optimism Bias and Strategic Misrepresentation in Planning:
Reference Class Forecasting in Practice." European Planning Studies, vol. 16, no. 1, January 2008, pp. 3-21.

"Truth and Lies About Megaprojects." Faculty of Technology, Policy, and Management, Delft University of Technology, September 2007.

"Policy and Planning for Large-Infrastructure Projects: Problems, Causes, Cures." Environment and Planning B: Planning and Design, vol. 34, 2007, pp. 578-597. This article was awarded the Association of European Schools of Planning (AESOP) Prize for Best Published Paper, July 2008.

"Planning and Design of Large Infrastructure Projects." Guest editorial. Co-author: Hugo Priemus. Environment and Planning B: Planning and Design, vol. 34, 2007, pp. 576-577.

"How Optimism Bias and Strategic Misrepresentation in Early Project Development Undermine Implementation." In Kjell J. Sunnevåg, ed., Beslutninger på svakt informasjonsgrunnlag: Tilnærminger og utfordringer i projekters tidlige fase (Decisions based on weak information: Approaches and challenges in the early phase of projects) (Trondheim, Norway: Concept Program, The Norwegian University of Science and Technology, 2007), pp. 41-55.

"Good Practice Lessons from the Urban Traffic Project, Denmark." Invited paper presented at Conference on Good Practice in Integration of Environment into Transport Policy, European Commission, Brussels, Belgium, 10-11 October 2002, 7 pp.

"Response to Phil Hodkinson." The British Journal of Educational Psychology, vol. 72, part 3, September 2002, pp. 452-53.

"Underestimating Costs in Public Works Projects: Error or Lie?" Principal author: Bent Flyvbjerg; co-authors: Mette K. Skamris Holm and Søren L. Buhl. Journal of the American Planning Association, vol. 68, no. 3, Summer 2002, pp. 279-295.

"Bringing Power to Planning Research: One Researcher's Praxis Story." Journal of Planning Education and Research, vol. 21, no. 4, Summer 2002, pp. 353-366.

"Planning and Foucault: In Search of the Dark Side of Planning Theory." Co-author: Tim Richardson. In Philip Allmendinger and Mark Tewdwr-Jones, eds., Planning Futures: New Directions for Planning Theory. London and New York: Routledge, 2002, pp. 44-62.

"Big Decisions, Big Risks: Improving Accountability in Mega Projects." Co-authors: Nils Bruzelius and Werner Rothengatter. Transport Policy, vol. 9, no. 2, April 2002, pp. 143-154.

"Bringing Power to Planning Research: One Researcher's Story." Keynote paper for the conference Planning Research 2000, London School of Economics and Political Science, 27-29 March 2000; published in Andy Thornley and Yvonne Rydin, eds. Planning in a Global Era. Aldershot: Ashgate, 2002, pp. 117-141.

Making Social Science Matter: Why Social Inquiry Fails and How It Can Succeed Again. Translated by Steven Sampson. Cambridge: Cambridge University Press, 2001, 10th printing 2007, 214 pp.

"Beyond the Limits of Planning Theory: Response to my Critics." International Planning Studies, vol. 6, no. 3, August 2001, pp. 285-292.

"The Science Wars." Philosophy and Social Action, vol. 27, no. 2, 2001, pp. 11-15.

"Ideal Theory, Real Rationality: Habermas Versus Foucault and Nietzsche." Paper for the Political Studies Association's 50 th Annual Conference, " The Challenges for Democracy in the 21 st Century," London School of Economics and Political Science, 10-13 April 2000, 20 pp.

"The Power of Rationality is Embedded in Stable Power Relations Rather Than in Confrontations." Philosophy and Social Action, vol. 25, no. 2, April-June 1999, pp. 37-41.

Rationality and Power: Democracy in Practice. Translated by Steven Sampson. University of Chicago Press, 1998, 5th printing 2005, 304 pp.

International Development Agency and Philippine National Economic and Development Authority, July 1988, 115 pp.

Handbook for Feasibility Studies and Project Planning. Ed. Copenhagen and Manila: Danish International Development Agency and Philippine National Economic and Development Authority, 1988, 290 pp.

"The Project of Planning: An Interview with John Friedmann." Scandinavian Housing and Planning Research, vol. 3, no. 2, 1986, pp. 103-117.

"Evaluation of Public Transport: Method for Application in Open Planning." Co-authors: Kjeld Kahr, Peter Bo Pedersen, and Johs. Vibe-Petersen. Transportation, vol. 13, no. 1, 1986, pp. 23-52.

"Citizen Participation Under Attack." Introductory article to special issue on democracy and citizen participation. Colloqui: Cornell Journal of Planning and Urban Issues, vol. 1, no. 1, 1986, pp. 5-6.

"Implementation and the Choice of Evaluation Methods." Transport Policy and Decision Making, special issue on policy implementation, vol. 2, no. 3, 1984, pp. 291-314.

"The Open Format and Citizen Participation in Transportation Planning." In Transportation Research Board, Social and Technological Issues in Transportation Planning. Transportation Research Record, no. 991. Washington, D.C.: National Research Council, 1984, pp. 15-22.

Planning Theory: Theoretical Considerations on the Analysis of Public Policy and Planning. Co-author: Verner C. Petersen. Aalborg University Press, 1983, 69 pp.

"Planning in the 33 Years after 1984." Co-author: Verner C. Petersen. In Patsy Healey et al., eds., Planning Theory: Prospects for the 1980's. Oxford: Pergamon Press, 1982, pp. 23-42.

"A Materialistic Concept of Planning and Participation." Co-author: Verner C. Petersen. Acta Sociologica, vol. 24, no. 4, 1981, pp. 293-311.

Mental Maps of Danish Schoolchildren. Co-authors: Erik Christiansen, Chris Jensen-Butler, Susanne Jeppesen, and Poul Tang Sørensen. Working Paper no. 6. Aarhus: Department of Geography, Aarhus University, 1978, 42 pp.

# Advisory Roles for DfT and HMT

Advisory roles for DfT and HMT have included the following
- Guidance document on optimism bias procedures (2004)
- Cost risk exposure assurance for HS2 phase 1 (2014-15)
- Cost risk exposure and optimism bias evaluation for HS2 phase 2B (2016)
- Cost risk exposure and optimism bias evaluation for the Midlands Mainline Project (2016)

## Conflicts of Interest with Regards to the Edinburgh Tram Project

No conflicts of interest are known. The authors of this report have not worked on the Edinburgh Tram project prior to the inquiry.